\documentclass{JHEP}	
\usepackage{epsfig}

\newcommand{\eref}[1]{(\ref{#1})}

\newcommand{\fref}[1]{Figure~\ref{#1}}
\newcommand{\cref}[1]{Chapter~\ref{#1}}
\newcommand{\beq}{\begin{equation}}
\newcommand{\eeq}{\end{equation}}
\newcommand{\ba}{\begin{array}}
\newcommand{\ea}{\end{array}}
\newcommand{\bcenter}{\begin{center}}
\newcommand{\ecenter}{\end{center}}

\def\IB{\relax\hbox{$\inbar\kern-.3em{\rm B}$}}
\def\IC{\relax\hbox{$\inbar\kern-.3em{\rm C}$}}
\def\ID{\relax\hbox{$\inbar\kern-.3em{\rm D}$}}
\def\IE{\relax\hbox{$\inbar\kern-.3em{\rm E}$}}
\def\IF{\relax\hbox{$\inbar\kern-.3em{\rm F}$}}
\def\IG{\relax\hbox{$\inbar\kern-.3em{\rm G}$}}
\def\IGa{\relax\hbox{${\rm I}\kern-.18em\Gamma$}}
\def\IH{\relax{\rm I\kern-.18em H}}
\def\IK{\relax{\rm I\kern-.18em K}}
\def\IL{\relax{\rm I\kern-.18em L}}
\def\IP{\relax{\rm I\kern-.18em P}}
\def\IR{\relax{\rm I\kern-.18em R}}
\def\IZ{\relax\ifmmode\mathchoice
{\hbox{\cmss Z\kern-.4em Z}}{\hbox{\cmss Z\kern-.4em Z}}
{\lower.9pt\hbox{\cmsss Z\kern-.4em Z}}
{\lower1.2pt\hbox{\cmsss Z\kern-.4em Z}}\else{\cmss Z\kern-.4em Z}\fi}
\def\II{\relax{\rm I\kern-.18em I}}

\def\sCC{{\kern 0.27em\vrule height1.45ex width0.03em depth0em
          \kern-0.30em\rm C}}
\def\C{{\mathchoice
  {\sCC}
  {\sCC}
  {\kern 0.225em \vrule height1.05ex width0.025em depth0em \kern-0.25em \rm C}
  {\kern 0.180em \vrule height0.78ex width0.02em depth0em \kern-0.2em \rm C}
        }}
\def\sHH{{\rm I\kern-.16em{}H}}
\def\H{{\mathchoice
  {\sHH}
  {\sHH}
  {\rm I\kern-.13em{}H}
  {\rm I\kern-.13em{}H} }}
\def\sNN{{\rm I\kern-.16em{}N}}
\def\N{{\mathchoice
  {\sNN}
  {\sNN}
  {\rm I\kern-.12em{}N}
  {\rm I\kern-.10em{}N} }}
\def\sPP{{\rm I\kern-.16em{}P}}
\def\P{{\mathchoice
  {\sPP}
  {\sPP}
  {\rm I\kern-.12em{}P}
  {\rm I\kern-.10em{}P} }}
\def\sQQ{{\kern 0.27em \vrule height1.45ex width0.03em depth0em
          \kern-0.30em \rm Q}}
\def\Q{{\mathchoice
        {\sQQ}
        {\sQQ}
  {\kern 0.225em \vrule height1.05ex width0.025em depth0em \kern-0.25em \rm Q}
  {\kern 0.180em \vrule height0.78ex width0.020em depth0em \kern-0.20em \rm Q}
        }}
\def\sRR{{\rm I\kern-0.16em{}R}}
\def\R{{\mathchoice
  {\sRR}
  {\sRR}
  {\rm I\kern-0.12em{}R}
  {\rm I\kern-0.10em{}R} }}
\def\sZZ{{\rm Z\kern-0.32em{}Z}}
\def\Z{{\mathchoice
  {\sZZ}
  {\sZZ} 
  {\rm Z\kern-0.3em{}Z}     
  {\rm Z\kern-0.25em{}Z} }}  
\def\ZZZ{{\rm Z\kern-0.24em{}Z}}
\def\sII{{\rm I\kern-0.16em{}I}}
\def\I{{\mathchoice
  {\sII}
  {\sII}
  {\rm I\kern-0.12em{}I}
  {\rm I\kern-0.10em{}I} }}

\def\dim{{\rm dim}}

\def\inbar{\,\vrule height1.5ex width.4pt depth0pt}
\font\cmss=cmss10 \font\cmsss=cmss10 at 7pt

\def\smiley{\hbox{\large$\bigcirc$\hspace{-0.80em}\raise.2ex
\hbox{$\cdot\cdot$}\kern-.61em\lower.2ex\hbox{\scriptsize$\smile$}}\ }
\def\frowny{\hbox{\large$\bigcirc$\hspace{-0.80em}\raise.2ex
\hbox{$\cdot\cdot$}\kern-.635em\lower.2ex\hbox{\scriptsize$\frown$}}\ }

\def\I{{\rlap{1} \hskip 1.6pt \hbox{1}}}
\def\sq{\framebox(10,10){~}}
\newcommand{\gen}[1]{\langle #1 \rangle}

\makeatletter
\let\hangafter\@hangfrom
\makeatother

\newtheorem{definition}{\sf DEFINITION}

\newtheorem{proposition}{\sf PROPOSITION}

\newtheorem{corollary}{\sf COROLLARY}

\newcommand{\bee}[1]{\left(\begin{array}{cc} i^{#1} & 0 \\ 0 & i^{#1} \\
	\end{array} \right)}

\preprint{MIT-CTP-3054\\ \\ {\tt hep-th/}}

\title{Stepwise Projection: Toward Brane Setups for Generic Orbifold
	Singularities}
\author{Bo Feng, Amihay Hanany, Yang-Hui He and Nikolaos Prezas
\footnote{
Research supported in part
by the Reed Fund Award,
the CTP and the LNS of MIT and the U.S. Department of Energy 
under cooperative research agreement \# DE-FC02-94ER40818.
A. H. is also supported by an A. P. Sloan Foundation Fellowship and 
a DOE OJI award.}
\\
Center for Theoretical Physics,
\\ Massachusetts Institute of Technology,\\ Cambridge, MA 02139, USA.\\
\email{fengb, hanany, yhe, prezas@ctp.mit.edu}
}
\abstract{The construction of brane setups for the
exceptional series $E_{6,7,8}$ 
of $SU(2)$ orbifolds remains an ever-haunting
conundrum. Motivated by techniques in some works by Muto  
on non-Abelian $SU(3)$ orbifolds, we here provide an algorithmic
outlook, a method which we call stepwise projection, 
that may shed some light on this puzzle. We exemplify this method,
consisting of transformation rules for obtaining complex quivers and
brane setups
from more elementary ones, to the cases of the $D$-series and $E_6$
finite subgroups of $SU(2)$. Furthermore, we demonstrate the
generality of the stepwise procedure by appealing to Fr{\o}benius'
theory of Induced Representations.
Our algorithm suggests the existence of
generalisations of the orientifold plane in string theory.}
\begin{document}
\section{Introduction}
It is by now a well-known fact that a stack of $n$ parallel coincident
D3-branes has on its world-volume, an ${\cal N}=4$, four-dimensional
supersymmetric $U(n)$ gauge theory. Placing such a stack at an orbifold
singularity of the form $\C^k/\{ \Gamma \subset SU(k) \}$ reduces
the supersymmetry to ${\cal N}=2,1$ and 0, respectively
for $k=2,3$ and $4$, and the gauge group is broken down to a product
of $U(n_i)$'s \cite{DM,JM,LNV}.

Alternatively, one could realize the gauge theory living on D-branes
by the so-called Brane Setups \cite{Han-Wit,Giveon} (or ``Comic
Strips'' as dubbed by Rabinovici \cite{Comic}) where D-branes are
stretched between NS5-branes and orientifold planes. Since these two
methods of orbifold projections and brane setups provide the same gauge
theory living on D-branes, there should exist some kind of duality to
explain the connection between them.

Indeed, we know now that by
T-duality one can map D-branes probing certain classes of orbifolds to 
brane configurations. For example, the two-dimensional orbifold
$\C^2/\{\IZ_k \subset SU(2) \}$, 
also known as an ALE singularity of type $A_{k-1}$, is mapped into a circle
of $k$ NS-branes (the so-called elliptic model) after proper T-duality
transformations. Such a mapping is easily generalized to some other
cases, such as the three-dimensional orbifold $\C^3/\{ \IZ_k\times \IZ_l
\subset SU(3) \}$ being mapped to the so-named Brane Box 
Model \cite{Han-Zaf,Han-Ura} or the four-dimensional case of $\C^4/\{ \IZ_k \times
\IZ_l \times \IZ_m\subset SU(4) \}$ being mapped to the brane cube model
\cite{BCM}. With the help of orientifold planes, we can T-dualise
$\C^2/\{ D_k \subset SU(2) \}$ to a brane configuration with
$ON$-planes \cite{Sen,Kapustin}, or $\C^3/\{ \IZ_k \times D_l
\subset SU(3) \}$ to brane-box-like models with $ON$-planes
\cite{Bo-Han}.

A further step was undertaken by Muto \cite{Muto,Muto2,Muto3} where an
attempt was made to establish the brane setup which corresponds to the
three-dimensional non-Abelian orbifolds $\C^3/\{ \Gamma \subset SU(3)
\}$ with $\Gamma=\Delta(3n^2)$ and $\Delta(6n^2)$. The key idea was
to arrive at these theories by judiciously quotienting the well-known
orbifold $\C^3/\{ \IZ_k\times \IZ_l \subset SU(3) \}$ whose brane
configuration is the Brane Box Model. 
In the process of this quotienting, a non-trivial $\IZ_3$ action on the
brane box is required. Though mathematically obtaining the quivers of
the former from those of the latter seems perfectly sound, such a
$\IZ_3$ action appears to be an unfamiliar symmetry in string theory. 
We shall briefly address this point later.

Now, with the exception of the above list of examples, there have been
no other successful brane setups for the myriad of orbifolds in
dimension two, three and four. Since we believe that the 
methods of orbifold projection and brane configurations are
equivalent to each other in giving D-brane world-volume gauge
theories, finding the T-duality mappings for arbitrary orbifolds is of 
great interest.

The present work is a small step toward such an aim. In particular, we
will present a so-called {\bf stepwise projection} algorithm which
attempts to systematize the quotienting idea of Muto, and, as we hope,
to give hints on the brane construction of generic orbifolds.

We shall chiefly focus on the orbifold projections by the $SU(2)$
discrete subgroups $D_k$ and $E_6$ in relation to $\IZ_n$. Thereafter,
we shall
evoke some theorems on induced representations which justify why our
algorithm of stepwise 
projection should at least work in general mathematically.
In particular, we will first demonstrate
how the algorithm gives the quiver of $D_k$ from that of $Z_{2k}$. We
then interpret this mathematical projection physically as precisely
the orientifold projection, whereby arriving at the brane setup of $D_k$ from
that of $\IZ_{2k}$, both of which are well-known and hence giving us a
consistency check.

Next we apply the same idea to $E_6$. We find that one can
construct its quiver from that of $\IZ_6$ or $D_2$ by an appropriate
$\IZ_3$ action. This is slightly mysterious to us physically as it
requires a $\IZ_3$ symmetry in string theory which we could use to
quotient out the $\IZ_6$ brane setup; such a symmetry we do not
know at this moment. However, in comparison with Muto's work, our
$\IZ_3$ action and the $\IZ_3$ investigated by Muto in light of the
$\Delta$ series of $SU(3)$, hint that there might be some objects in
string theory which provide a $\IZ_3$ action, analogous to the orientifold
giving a $\IZ_2$, and which we could use on the known brane setups to
establish those yet unknown, such as those corresponding to the
orbifolds of the exceptional series.

The organisation of the paper is as follows. In \S 2 we review the
technique of orbifold projections in an explicit matrix language
before moving on to \S 3 to present our stepwise projection
algorithm. In particular, \S 3.1 will demonstrate how to obtain the
$D_k$ quiver from the $\IZ_{2k}$ quiver, \S 3.2 and \S 3.3 will show how
to get that of $E_6$ from those of $D_2$ and $\IZ_6$ respectively. We
finish with comments on the algorithm in \S 4. We will use induced
representation theory in \S 4.1 to prove the validity of our methods
and in \S 4.2 we will address how the present work may be used as a
step toward the illustrious goal of obtaining brane setups for the
generic orbifold singularity.

During the preparation of the manuscript, it has come to our attention
that independent and variant forms of the method have been in
germination \cite{Uranga,Berenstein}; we sincerely hope that our
systematic treatment of the procedure may be of some utility thereto.

\section*{Nomenclature}

Unless otherwise stated we shall adhere to the convention that
$\Gamma$ refers to a discrete subgroup of $SU(n)$ (i.e., a finite
collineation group), that $\gen{x_1,..,x_n}$ is a finite group generated
by $\{x_1,..,x_n\}$, that $|\Gamma|$ is the order of the group
$\Gamma$, that $D_k$ is the binary dihedral group of order $4k$,
that $E_{6,7,8}$ are the binary exceptional subgroups of $SU(2)$, and that
$R_{G(n)}^{\bullet}(x)$ is a representation of the element $x \in G$ of
dimension $n$ with $\bullet$ denoting properties such as regularity,
irreducibility, etc., and/or simply a label. Moreover, $S^T$ shall denote
the transpose of the matrix $S$ and $A \otimes B$ is the tensor
product of matrices $A$ and $B$ with block matrix elements $A_{ij} B$.
Finally we frequently use the Pauli matrices $\{\sigma_i, i=1,2,3\}$
as well as $\I_N$ for the $N \times N$ identity matrix.
We emphasise here that
the notation for the binary groups differs from our other works
in the exclusion of~$\widehat{~}$~and in the convention for the
sub-index of the binary dihedral group.
\section{A Review on Orbifold Projections}
The general methodology of how the finite group structure of the
orbifold projects the gauge theory has been formulated in
\cite{LNV}. The complete lists of two and three dimensional cases have
been treated respectively in \cite{DM,JM} and \cite{Han-He,Muto} as
well as the four dimensional case in \cite{4d}.
For the sake of our forth-coming discussion, we shall not use the
nomenclature in \cite{LNV,Han-He,Bo-Han} where recourse to McKay's
Theorem and abstractions to representation theory are taken. Instead,
we shall adhere to the notations in \cite{JM} and explicitly indicate
what physical fields survive the orbifold projection.

Throughout we shall focus on two dimensional orbifolds $\C^2/\{\Gamma
\subset SU(2)\}$. The parent theory has an $SU(4)\cong Spin(6)$
R-symmetry from the ${\cal N}=4$ SUSY. The $U(n)$ gauge bosons $A^\mu_{IJ}$
with $I,J=1,...,n$ are R-singlets. Furthermore, there are Weyl
fermions $\Psi_{IJ}^{i=1,2,3,4}$ in the fundamental $\bf{4}$ of $SU(4)$ and
scalars $\Phi_{IJ}^{i=1,..,6}$ in the antisymmetric $\bf{6}$.

The orbifold imposes a projection condition upon these fields due to
the finite group $\Gamma$. Let $R^{reg}_\Gamma(g)$ be the regular
representation of $g \in \Gamma$, by which we mean
\[
R^{reg}_{\Gamma}(g) := \bigoplus\limits_i \Gamma_i(g) \otimes
	\I_{\dim(\Gamma_i)}
\]
where $\{\Gamma_i\}$ are the irreducible representations of
$\Gamma$. In matrix form, $R^{reg}_{\Gamma}(g)$ is composed of blocks of irreps,
with each of dimension $j$ repeated $j$ times. Therefore it is a
matrix of size $\sum\limits_i \dim(\Gamma_i)^2 = |\Gamma|$.
Let Irreps$(\Gamma) = \{\Gamma_1^{(1)},\ldots,\Gamma_{m_1}^{(1)};
\Gamma_1^{(2)},\ldots,\Gamma_{m_2}^{(2)};\ldots \ldots;
\Gamma_{1}^{(n)},\ldots,\Gamma_{m_n}^{(n)}\}$, 
consisting of $m_j$ irreps of dimension $j$, then
\beq
\label{reg}
\hspace{-0.7in}
R^{reg}_\Gamma := 
{\tiny
\left(
\ba{cccccccccc}
\Gamma_1^{(1)} &  &  &  &  &  &  &  &  &  \\
 & \ddots &  &  &  &  &  &  &  &  \\
 &  & \Gamma_{m_1}^{(1)} &  &  &  &  &  &  &  \\
	 &  &  & \left(\matrix{\Gamma_1^{(2)} & ~ \cr
	~ & \Gamma_1^{(2)}}\right) &  &  &  &  &  &  \\
 &  &  &  & \ddots &  &  &  &  &  \\
 &  &  &  &  & \left(\matrix{\Gamma_{m_2}^{(2)} & ~ \cr
	~ & \Gamma_{m_2}^{(2)}}\right) &  &  &  &  \\
 &  &  &  &  &  & \ddots &  &  &  \\
 &  &  &  &  &  &  &
	\left(\ba{ccc}
	\Gamma_1^{(n)} & & \\
	& \ddots & \\
	& & \Gamma_1^{(n)}
	\ea\right)_{n \times n} &  &  \\
 &  &  &  &  &  &  &  & \ddots &  \\
 &  &  &  &  &  &  &  &  & 
	\left(\ba{ccc}
	\Gamma_{m_n}^{(n)} & & \\
	& \ddots & \\
	& & \Gamma_{m_n}^{(n)} \\
	\ea\right)_{n \times n}
\ea
\right)
}.
\eeq
Of the parent fields $A^\mu, \Psi, \Phi$, only those invariant under
the group action will remain in the orbifolded theory; this imposition
is what we mean by {\em surviving the projection}:
\beq
\label{proj}
\ba{l}
A^\mu = R^{reg}_{\Gamma}(g)^{-1} \cdot A^\mu \cdot R^{reg}_{\Gamma}(g)\\
\Psi^i = \rho(g)^i_j ~R^{reg}_{\Gamma}(g)^{-1} \cdot \Psi^j 
	\cdot R^{reg}_{\Gamma}(g)\\
\Phi^i = \rho'(g)^i_j ~R^{reg}_{\Gamma}(g)^{-1} \cdot \Phi^j \cdot
	R^{reg}_{\Gamma}(g) 
\qquad
\forall~~g\in\Gamma,
\ea
\eeq
where $\rho$ and $\rho'$ are induced actions because the matter fields
carry R-charge (while the gauge bosons are R-singlets).
Clearly if $\Gamma = \gen{x_1, ...,  x_n}$, it suffices to
impose \eref{proj} for the generators $\{x_i\}$ in order to find the
matter content of the orbifold gauge theory; this observation we shall
liberally use henceforth.

Letting $n = N |\Gamma|$ for some large $N$ and $n_i = \dim(\Gamma_i)$, 
the subsequent gauge group becomes $\prod\limits_i U(n_iN)$ with
$a_{ij}^4$ Weyl fermions as bifundamentals
$\left({\bf n_iN},{\bf \overline{n_jN}}\right)$ as well as $a_{ij}^6$ scalar
bifundamentals. These bifundamentals are pictorially summarised in
quiver diagrams whose adjacency matrices are the $a_{ij}$'s.

Since we shall henceforth be dealing primarily with $\C^2$
orbifolds, we have ${\cal N}=2$ gauge theory in four dimensions
\cite{LNV}. In particular we choose the induced group action on the
R-symmetry to be ${\bf 4} = {\bf1}_{trivial}^2 \oplus {\bf 2}$ and
${\bf 6} = 
{\bf 1}_{trivial}^2 \oplus {\bf 2}^2$ in order to preserve the
supersymmetry. For 
this reason we can specify the final fermion and scalar matter
matrices by a single quiver characterised by the ${\bf 2}$ of $SU(2)$
as the trivial ${\bf 1}$'s give diagonal 1's. These issues are 
addressed at length in \cite{Han-He}.
\section{Stepwise Projection}
Equipped with the clarification of notations of the previous section
we shall now illustrate a technique which we shall call {\bf stepwise
projection}, originally inspired
by \cite{Muto,Muto2,Muto3}, who attempted brane realisations of certain
non-Abelian orbifolds of $\C^3$, an issue to which we shall later turn.

The philosophy of the technique is straight-forward\footnote{A recent
work \cite{Berenstein} appeared during
the final preparations of this draft; it beautifully
addresses issues along a similar vein. In particular, cases where
$\Gamma_1$ is normal in $\Gamma_2$ are discussed in detail.
However, our stepwise method is not restricted by normality.}:
say we are given a group $\Gamma_1 = \gen{x_1,...,x_n}$ with
quiver diagram $Q_1$ and $\Gamma_2 = \gen{x_1,...,x_{n+1}} \supset
\Gamma_1$  with quiver $Q_2$, we wish to determine $Q_2$ from $Q_1$ by
the projection \eref{proj} by $\{x_1,...,x_n\}$ followed by another
projection by $x_{n+1}$.

We now proceed to analyse the well-known examples of the
cyclic and binary dihedral quivers under this new light.
\subsection{$D_k$ Quivers from $A_k$ Quivers}
We shall concern ourselves with orbifold theories of $\C^2/\IZ_k$ and
$\C^2/D_k$.
Let us first recall that the cyclic group $A_{k-1} \cong \IZ_k$
has a single generator
\[
\beta_k := \left(
\matrix{\omega_k & 0 \cr 0 & \omega_k^{-1}} \right),
\qquad \mbox{with} \quad \omega_n := e^{2 \pi i \over n}
\]
and that the generators for the binary dihedral group $D_{k}$ are
\[
\beta_{2k} = \left(
\matrix{\omega_{2k} & 0 \cr 0 & \omega_{2k}^{-1}} \right),
\qquad
\gamma := \left( \matrix{0 & i \cr i & 0} \right).
\]
We further recall from \cite{Bo-Han} that 
$D_k / \IZ_{2k} \cong \IZ_2$.

Now all irreps for $\IZ_k$ are 1-dimensional (the $k^{th}$ roots of
unity), and \eref{reg} for the generator reads
\[
R^{reg}_{\IZ_k}(\beta_k) = \left(
\begin{array}{ccccc}
1 & 0 & 0 & 0 & 0 \\
0 & \omega_k & 0 & 0 & 0 \\
0 & 0 & \omega^2_k & 0 & 0 \\
0 & 0 & 0 & \ddots & 0 \\
0 & 0 & 0 & 0 & \omega_k^{k-1}
\end{array}
\right).
\]
On the other hand, $D_{k}$ has 1 and 2-dimensional irreps and
\eref{reg} for the two generators become
\[
\hspace{-1.0in}
R^{reg}_{D_k}(\beta_{2k}) = 
{\tiny
\left(
\begin{array}{ccccccc}
\left( \matrix{1 & 0 \cr 0 & -1} \right) & 0 & 0 & 0 & 0 & 0 & 0 \\
0 & \left( \matrix{1 & 0 \cr 0 & -1} \right) & 0 & 0 & 0 & 0 & 0 \\
0 & 0 & \left( \matrix{\omega_{2k} & 0 \cr 0 & \omega_{2k}^{-1}}
	\right) & 0 & 0 & 0 & 0 \\
0 & 0 & 0 & \left( \matrix{\omega_{2k} & 0 \cr 0 & \omega_{2k}^{-1}}
	\right) & 0 & 0 & 0\\
\vdots & \vdots & \vdots & \vdots & \ddots & \vdots & \vdots \\
0 & 0 & 0 & 0 & 0 & 
	\left( \matrix{\omega^{k-1}_{2k} & 0 \cr 0 & \omega_{2k}^{-(k-1)}}
	\right)& 0 \\
0 & 0 & 0 & 0 & 0 & 0 &
	\left( \matrix{\omega^{k-1}_{2k} & 0 \cr 0 & \omega_{2k}^{-(k-1)}}
	\right)
\end{array}
\right)
}
\]
and
\[
\hspace{-1.0in}
R^{reg}_{D_k}(\gamma) = 
{\tiny
\left(
\begin{array}{ccccccc}
\left(
\matrix{1 & 0 \cr 0 & i^{k \bmod 2}} \right) & 0 & 0 & 
0 & 0 & 0 & 0 \\ 
0 & \left( \matrix{-1 & 0 \cr 0 & -i^{k \bmod 2}} \right)
& 0 & 0 & 0 & 0 & 0 \\
0 & 0 & \left( \matrix{0 & i \cr i & 0} \right) & 0 & 0 & 0 & 0 \\
0 & 0 & 0 & \left( \matrix{0 & i \cr i & 0} \right) & 0 & 0 & 0\\
\vdots & \vdots & \vdots & \vdots & \ddots & \vdots & \vdots \\
0 & 0 & 0 & 0 & 0 & \left( \matrix{0 & i^{k-1} \cr i^{k-1} &
	0}\right)& 0 \\ 
0 & 0 & 0 & 0 & 0 & 0 & \left( \matrix{0 & i^{k-1} \cr i^{k-1} &
	0}\right)
\end{array}
\right)
}.
\]
In order to see the structural similarities between the regular
representation of $\beta_{2k}$ in $\Gamma_1 = \IZ_{2k}$ and $\Gamma_2 =
D_k$, we need to perform a change of basis. We do so such that each
pair (say the $j^{th}$) of the 2-dimensional irreps of $D_2$ becomes
as follows:
\[
\Gamma^{(2)}(\beta_{2k}) =
\left(
\begin{array}{cc}
\left( \matrix{\omega^{j}_{2k} & 0 \cr 0 & \omega_{2k}^{-j}} \right) &
	0 \\
0 & \left( \matrix{\omega^{j}_{2k} & 0 \cr 0 & \omega_{2k}^{-j}}
	\right)
\end{array}
\right)
\rightarrow
\left(
\begin{array}{cc}
\omega^{j}_{2k} \left( \matrix{1 & 0 \cr 0 & 1} \right) & 0 \\
0 & \omega_{2k}^{-j} \left( \matrix{1 & 0 \cr 0 & 1} \right)
\end{array}
\right)
\]
where $j = 1, 2, \ldots, k - 1.$
In this basis, the 2-dimensionals of $\gamma$ become
\[
\Gamma^{(2)}(\gamma) =
\left(
\begin{array}{cc}
\left( \matrix{0 & i^j \cr i^j & 0} \right) & 0 \\
0 & \left( \matrix{0 & i^j \cr i^j & 0} \right)
\end{array}
\right)
\rightarrow
\left(
\begin{array}{cc}
0 & i^j \left(\matrix{1 & 0 \cr 0 & 1}\right) \\
i^j \left(\matrix{1 & 0 \cr 0 & 1}\right) & 0
\end{array}
\right).
\]

Now for the 1-dimensionals, we also permute the basis:
{\scriptsize
\[
\hspace{-0.7in}
\Gamma^{(1)}(\beta_{2k}) =
\left(
\matrix{ 1 & 0 & 0 & 0 \cr 0 & 
     -1 & 0 & 0 \cr 0 & 0 & 1 & 0 \cr 0 & 0 & 0 & -1 \cr} 
\right)
\rightarrow
\left(
\matrix{ 1 & 0 & 0 & 0 \cr 0 & 1 & 0 & 0 \cr
	0 & 0 & -1 & 0 \cr 0 & 0 & 0 & -1 \cr  }
\right)
\qquad
\Gamma^{(1)}(\gamma) =
\left(
\matrix{ 1 & 0 & 0 & 0 \cr 0 & i^{k \bmod 2} & 0 & 0 \cr
	0 & 0 & -1 & 0 \cr 0 & 0 & 0 & -i^{k \bmod 2} \cr  }
\right)
\rightarrow
\left(
\matrix{ 1 & 0 & 0 & 0 \cr 0 & -1 & 0 & 0 \cr
	0 & 0 & i^{k \bmod 2} & 0 \cr
	0 & 0 & 0 & -i^{k \bmod 2} \cr  }
\right).
\]
}
Therefore, we have
{\scriptsize
\[
{\hspace{-0.7in}
R^{reg}_{D_k}(\beta_{2k}) =
\left(
\begin{array}{ccccccc}
1 & 0 & 0 & 0 &  & 0 & 0 \\
0 & -1 & 0 & 0 &  & 0 & 0 \\
0 & 0 & \omega_{2k} & 0 &  & 0 & 0 \\
0 & 0 & 0 & \omega_{2k}^{-1} &  & 0 & 0 \\
\vdots & \vdots & \vdots & \vdots & \ddots & \vdots & \vdots \\
0 & 0 & 0 & 0 & & \omega_{2k}^{k-1} & 0 \\
0 & 0 & 0 & 0 & & 0 & \omega_{2k}^{-(k-1)}
\end{array}
\right)
\otimes
\left( \matrix{1 & 0 \cr 0 & 1} \right),
}
\]
}
which by now has a great resemblance to the regular representation of
$\beta_{2k} \in \IZ_{2k}$; indeed, after one final change of basis, 
by ordering the powers of $\omega_{2k}$ in an ascending fashion while
writing $\omega_{2k}^{-j} = \omega_{2k}^{2k-j}$ to ensure only
positive exponents, we arrive at
\beq
\label{betafinDk}
\ba{rcl}
R^{reg}_{D_k}(\beta_{2k}) & = & 
\left(
\begin{array}{ccccc}
1 & 0 & 0 & & 0 \cr
0 & \omega_{2k} & 0 & & 0\cr
0 & 0 & \omega_{2k}^2 & & 0 \cr
\vdots & \vdots & \vdots & \ddots & \vdots \\
0 & 0 & 0 & & \omega_{2k}^{2k-1}
\end{array}
\right)
\bigotimes
\left( \matrix{1 & 0 \cr 0 & 1} \right)
\\
& = & R^{reg}_{\IZ_{2k}}(\beta_{2k}) \otimes \I_2,
\ea
\eeq
the key relation which we need.

Under this final change of basis,
{\scriptsize
\beq
\label{gammafin}
{\hspace{-0.55in}
R^{reg}_{D_k}(\gamma) =
\left(
\matrix{
\left( \matrix{1 & 0 \cr 0 & -1}\right)
	& 0 & 0 & 0 & 0 & 0 & 0 & 0 \cr
0 & 0 & 0 & 0 & 0 & \bee{k-3} & 0 & 0 \cr
0 & 0 & 0 & 0 & 0 & 0 & \bee{k-2} & 0 \cr
0 & \vdots & \vdots & \ddots & \vdots & \vdots & 0 & \bee{k-1} \cr 
\vdots & & & & \left(\matrix{i^{k \bmod 2} & 0 \cr 0 & -i^{k \bmod 2}}
	\right) & & & \vdots \cr
0 & \bee{k-3} & 0 & 0 & 0 & 0 & 0 & 0 \cr
0 & 0 & \bee{k-2} & 0 & 0 & 0 & 0 & 0 \cr
0 & 0 & 0 & \bee{k-1} & 0 & 0 & 0 & 0 \cr} 
\right)
}.
\eeq
}
Our strategy is now obvious. We shall first project according to
\eref{proj}, using \eref{betafinDk}, which is {\em equivalent to} a
projection by $\IZ_{2k}$, except with two identical copies (physically, this
simply means we place twice as many D3-brane probes). Thereafter we
shall project once again using \eref{gammafin} and the resulting theory
should be that of the $D_k$ orbifold.

\subsection*{An Illustrative Example}
Let us turn to a concrete example, namely $\IZ_4 \rightarrow D_2$. The
key points to note are that $D_2 :=\gen{\beta_4,\gamma}$
and $\IZ_4 \cong \gen{\beta_4}$. We shall therefore
perform stepwise projection by $\beta_4$ followed by $\gamma$.

Equation \eref{betafinDk} now reads
\beq
R^{reg}_{D_2}(\beta_4) = 
R^{reg}_{\IZ_{4}}(\beta_4) \otimes \I_2 = 
\left(\matrix{ 1 & 0 & 0 & 0 \cr 0 & i & 0 & 0 \cr
	0 & 0 & i^2 & 0 \cr 0 & 0 & 0 & i^3 \cr  } 
\right)
\otimes \I_2 .
\label{betafinD2}
\eeq
We have the following matter content in the parent (pre-orbifold)
theory: gauge field $A^\mu$, fermions $\Psi^{1,2,3,4}$ and scalars
$\Phi^{1,2,3,4,5,6}$ (suppressing gauge indices $IJ$).
Projection by $R^{reg}_{D_2}(\beta_4)$ in \eref{betafinD2} according
to \eref{proj} gives a $\IZ_4$ orbifold theory, which restricts the
form of the fields to be as follows:
\beq
\label{D2step1}
A^\mu, \Psi^{1,2}, \Phi^{1,2} =
\left(
\begin{array}{cccc}
\sq & & & \\ & \sq & & \\ & & \sq & \\ & & & \sq
\end{array}
\right);
\quad
\Psi^3, \Phi^{3,5} =
\left(
\begin{array}{cccc}
& \sq & & \\ & & \sq & \\ & & & \sq \\ \sq & & &
\end{array}
\right);
\quad
\Psi^4, \Phi^{4,6} =
\left(
\begin{array}{cccc}
& & & \sq \\ \sq & & & \\ & \sq & & \\ & & \sq &
\end{array}
\right)
\eeq
where $\sq$ are $2 \times 2$ blocks. We recall from the
previous section that we have chosen the R-symmetry decomposition as
${\bf 4} = {\bf 1}_{trivial}^2 \oplus {\bf 2}$ and
${\bf 6} = {\bf 1}_{trivial}^2 \oplus {\bf 2}^2$. The fields in
\eref{D2step1} are defined in accordance thereto: the fermions
$\Psi^{1,2}$ and scalars $\Phi^{1,2}$ are respectively in the two
trivial {\bf 1}'s of the {\bf 4} and {\bf 6}; $(\Psi^3,\Psi^4)$,
$(\Phi^3,\Phi^4)$ and $(\Phi^5,\Phi^6)$ are in the doublet {\bf 2} of $\Gamma$
inherited from $SU(2)$.
\EPSFIGURE[ht]{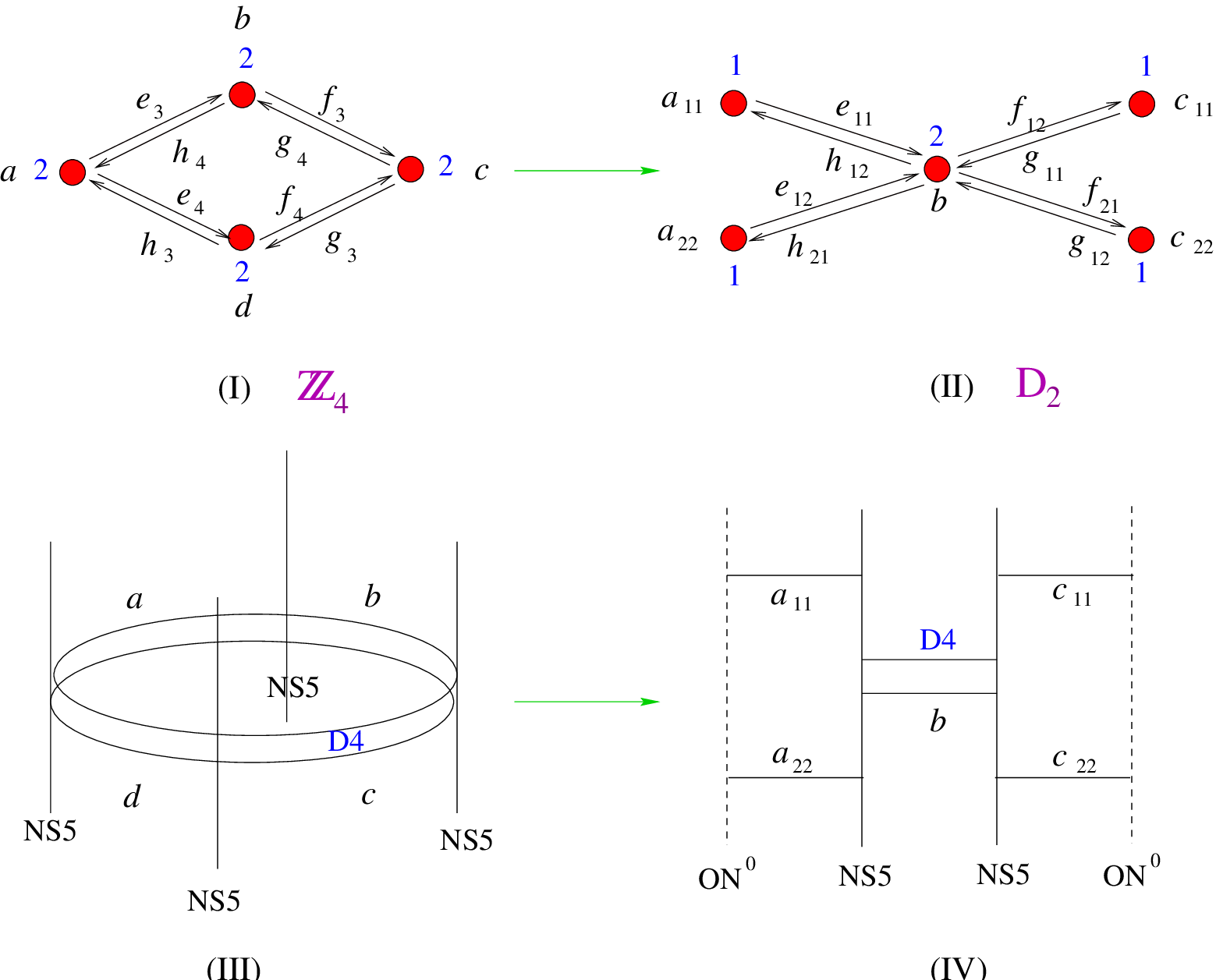,width=6.5in}
{
From the fact that $D_2 :=\gen{\beta_4,\gamma}$ is generated by $\IZ_4
= \beta_4$ together with $\gamma$, our stepwise projection, first by
$\beta_4$, and then by $\gamma$, gives 2 copies of the $\IZ_4$ quiver
in Part (I) and then the $D_2$ quiver in Part (II) by appropriate
joining/splitting of the nodes and arrows. The brane configurations
for these theories are given in Parts (III) and (IV).
\label{f:D2}
}
Indeed, the $R^{reg}_{\IZ_4}(\beta_4)$
projection would force $\sq$ to be numbers and not matrices as we do
not have the extra $\I_2$ tensored to the group action, in which case
\eref{D2step1} would be $4\times 4$ matrices prescribing the adjacency
matrices of the $\IZ_4$ quiver. For this
reason, the quiver diagram for the $\IZ_4$ theory as drawn in
part (I) of \fref{f:D2} has the nodes labelled 2's instead of the
usual Dynkin labels of 1's for the $A$-series. In physical terms we
have placed twice as many image D-brane probes. The key point is that because
$\sq$ are now matrices (and \eref{D2step1} are $8 \times 8$), further
projection internal thereto may change the number and structure of the
product gauge groups and matter fields.

Having done the first step by the $\beta_4$ projection,
next we project with the regular representation of $\gamma$:
\beq
\label{gammafin2}
R^{reg}_{D_2}(\gamma) = 
{\scriptsize
\left(
\begin{array}{cccc}
\left( \matrix{ 1 & 0 \cr 0 & -1 \cr  }\right) & 0 & 0 & 0 \cr
0 & 0 & 0 & \left( \matrix{ i & 0 \cr 0 & i \cr  }\right) \cr 
0 & 0 & \left( \matrix{ 1 & 0 \cr 0 & -1 \cr} \right) & 0 \cr
0 & \left( \matrix{ i & 0 \cr 0 & i \cr} \right) & 0 & 0 \cr
\end{array}
\right)
}
:=
\left(
\matrix{\sigma_3 & 0 & 0 & 0 \cr 0 & 0 & 0 & i\I_2 \cr
0 & 0 & \sigma_3 & 0 \cr 0 & i\I_2 & 0 & 0 \cr}
\right).
\eeq
In accordance with \eref{D2step1}, let the gauge field be
\[
A^\mu := \left(\matrix{a & 0 & 0 & 0 \cr 0 & b & 0 & 0 \cr
0 & 0 & c & 0 \cr 0 & 0 & 0 & d \cr} 
\right),
\]
with $a,b,c,d$ denoting the $2 \times 2$ blocks $\sq$, \eref{proj} for
\eref{gammafin2} now reads
\[
A^\mu = R^{reg}_{D_2}(\gamma)^{-1} \cdot A^\mu \cdot R^{reg}_{D_2}(\gamma)
\Rightarrow
\]
{\small
\[
\left(\matrix{a & 0 & 0 & 0 \cr 0 & b & 0 & 0 \cr
0 & 0 & c & 0 \cr 0 & 0 & 0 & d \cr} \right)
=
\left(
\matrix{\sigma_3 & 0 & 0 & 0 \cr 0 & 0 & 0 & -i\I_2 \cr
0 & 0 & \sigma_3 & 0 \cr 0 & -i\I_2 & 0 & 0 \cr}
\right)
\left(\matrix{a & 0 & 0 & 0 \cr 0 & b & 0 & 0 \cr
0 & 0 & c & 0 \cr 0 & 0 & 0 & d \cr} \right)
\left(
\matrix{\sigma_3 & 0 & 0 & 0 \cr 0 & 0 & 0 & i\I_2 \cr
0 & 0 & \sigma_3 & 0 \cr 0 & i\I_2 & 0 & 0 \cr}
\right),
\]
}
giving us a set of constraining equations for the blocks:
\beq
\label{D2cons1}
\sigma_3 \cdot a \cdot \sigma_3 = a; \qquad d = b; \qquad 
\sigma_3 \cdot c \cdot \sigma_3 = c.
\eeq
Similarly, for the fermions in the {\bf 2}, viz.,
\[
\Psi^3 = \left( \matrix{0 & e_3 & 0 & 0 \cr 0 & 0 & f_3 & 0\cr
0 & 0 & 0 & g_3 \cr h_3 & 0 & 0 & 0 \cr} \right),
\quad
\Psi^4 = \left( \matrix{0 & 0 & 0 & e_4 \cr f_4 & 0 & 0 & 0 \cr
0 & g_4 & 0 & 0 \cr 0 & 0 & h_4 & 0 \cr} \right),
\]
the projection \eref{proj} is
\[
\gamma \cdot \left( \begin{array}{cc} \Psi^3 \\ \Psi^4 \end{array}\right)
= R^{reg}_{D_2}(\gamma)^{-1} \cdot
 \left( \begin{array}{cc} \Psi^3 \\ \Psi^4 \end{array}\right)
 \cdot R^{reg}_{D_2}(\gamma).
\]
We have used the fact that the induced action $\rho(\gamma)$, having
to act upon a doublet, is simply the $2 \times 2$ matrix
$\gamma$ herself. Therefore, writing it out explicitly, we have
{\small
\[
i \left( \matrix{0 & 0 & 0 & e_4 \cr f_4 & 0 & 0 & 0 \cr
0 & g_4 & 0 & 0 \cr 0 & 0 & h_4 & 0 \cr} \right)
= \left(
\matrix{\sigma_3 & 0 & 0 & 0 \cr 0 & 0 & 0 & -i\I_2 \cr
0 & 0 & \sigma_3 & 0 \cr 0 & -i\I_2 & 0 & 0 \cr}
\right)
\left( \matrix{0 & e_3 & 0 & 0 \cr 0 & 0 & f_3 & 0\cr
0 & 0 & 0 & g_3 \cr h_3 & 0 & 0 & 0 \cr} \right)
\left(
\matrix{\sigma_3 & 0 & 0 & 0 \cr 0 & 0 & 0 & i\I_2 \cr
0 & 0 & \sigma_3 & 0 \cr 0 & i\I_2 & 0 & 0 \cr}
\right)
\]
}
and
{\small
\[
i \left( \matrix{0 & e_3 & 0 & 0 \cr 0 & 0 & f_3 & 0\cr
0 & 0 & 0 & g_3 \cr h_3 & 0 & 0 & 0 \cr} \right)
= \left(
\matrix{\sigma_3 & 0 & 0 & 0 \cr 0 & 0 & 0 & -i\I_2 \cr
0 & 0 & \sigma_3 & 0 \cr 0 & -i\I_2 & 0 & 0 \cr}
\right)
\left( \matrix{0 & 0 & 0 & e_4 \cr f_4 & 0 & 0 & 0 \cr
0 & g_4 & 0 & 0 \cr 0 & 0 & h_4 & 0 \cr} \right)
\left(
\matrix{\sigma_3 & 0 & 0 & 0 \cr 0 & 0 & 0 & i\I_2 \cr
0 & 0 & \sigma_3 & 0 \cr 0 & i\I_2 & 0 & 0 \cr}
\right),
\]
}
which gives the constraints
\beq
\label{D2cons2}
f_4 = -h_3 \cdot \sigma_3; \qquad g_4 = \sigma_3 \cdot g_3; \qquad
h_4 = -f_3 \cdot \sigma_3; \qquad e_4 = \sigma_3 \cdot e_3.
\eeq
The doublet scalars $(\Phi^{3,5},\Phi^{4,6})$ of course give the same
results, as should be expected from supersymmetry.

In summary then, the final fields which survive both $\beta_4$ and
$\gamma$ projections (and thus the entire group $D_2$) are
{\small
\beq
\label{D2step2}
\ba{l}
A^\mu = \left(
\begin{array}{cccc}
\left( \matrix{a_{11} & 0 \cr 0 & a_{22}} \right) & & &\\
& b & &\\
& & \left( \matrix{c_{11} & 0 \cr 0 & c_{22}} \right) & \\
& & & b
\end{array}
\right);
\quad
\left\{
\ba{c}
e_3 = \left( \matrix{e_{11} & e_{12} \cr 0 & 0} \right), \quad
f_3 = \left( \matrix{0 & f_{12} \cr 0 & f_{22}} \right), \\ \\
g_3 = \left( \matrix{g_{11} & g_{12} \cr 0 & 0} \right), \quad
h_3 = \left( \matrix{0 & h_{12} \cr 0 & h_{22}} \right),
\ea
\right.
\\
\Psi^3 = \left( \matrix{0 & e_3 & 0 & 0 \cr 0 & 0 & f_3 & 0\cr
0 & 0 & 0 & g_3 \cr h_3 & 0 & 0 & 0 \cr} \right),
\quad
\Psi^4 = \left(\matrix{0 & 0 & 0 & \sigma_3 \cdot e_3 \cr 
-h_3 \cdot \sigma_3 & 0 & 0 & 0 \cr
0 & \sigma_3 \cdot g_3 & 0 & 0 \cr
0 & 0 & -f_3 \cdot \sigma_3 & 0 \cr}
\right).
\ea
\eeq
}
The key features to be noticed are now apparent in the structure of
these matrices in \eref{D2step2}.
We see that the 4 blocks of $A^\mu$ in \eref{D2step1}, which give
the four nodes of the $\IZ_4$ quiver, now undergo a metamorphosis:
we have written out the components of $a,c$ explicitly and have used
\eref{D2cons1} to restrict both to diagonal matrices, while $b$ and
$d$ are identified, but still remain blocks without internal structure
of interest. Thus we have a total of 5 non-trivial constituents
$a_{11},a_{22},c_{11},c_{22}$ and $b$, precisely the 5
nodes of the $D_2$ quiver (see parts (I) and (II) of \fref{f:D2}).
Thus {\em nodes of the quiver merge and
split as we impose further projections}, as we mentioned a few
paragraphs ago.

As for the bifundamentals, i.e., the arrows of the quiver,
\eref{D2step1} prescribes the blocks $e_{3,4}, f_{3,4}, g_{3,4}$ and
$h_{3,4}$ as the 8 arrows of Part (I) of \fref{f:D2}. After the
projection by $\gamma$, and imposing the constraint \eref{D2cons2} as
well as the fact that all entries of matter matrices must be
non-negative, we are left with the 8 fields $e_{11,12}, f_{12,22},
g_{11,12}$ and $h_{12,22}$, precisely the 8 arrows in the $D_2$ quiver
(see Part (II) of \fref{f:D2}).

\subsection*{The General Case}
The generic situation of obtaining the $D_k$ quiver from that of
$\IZ_{2k}$ is completely analogous. We would always have two end nodes of
the $\IZ_{2k}$ quiver each splitting into two while the middle ones
coalesce pair-wise, as is shown in \fref{f:Dk}.
\EPSFIGURE[ht]{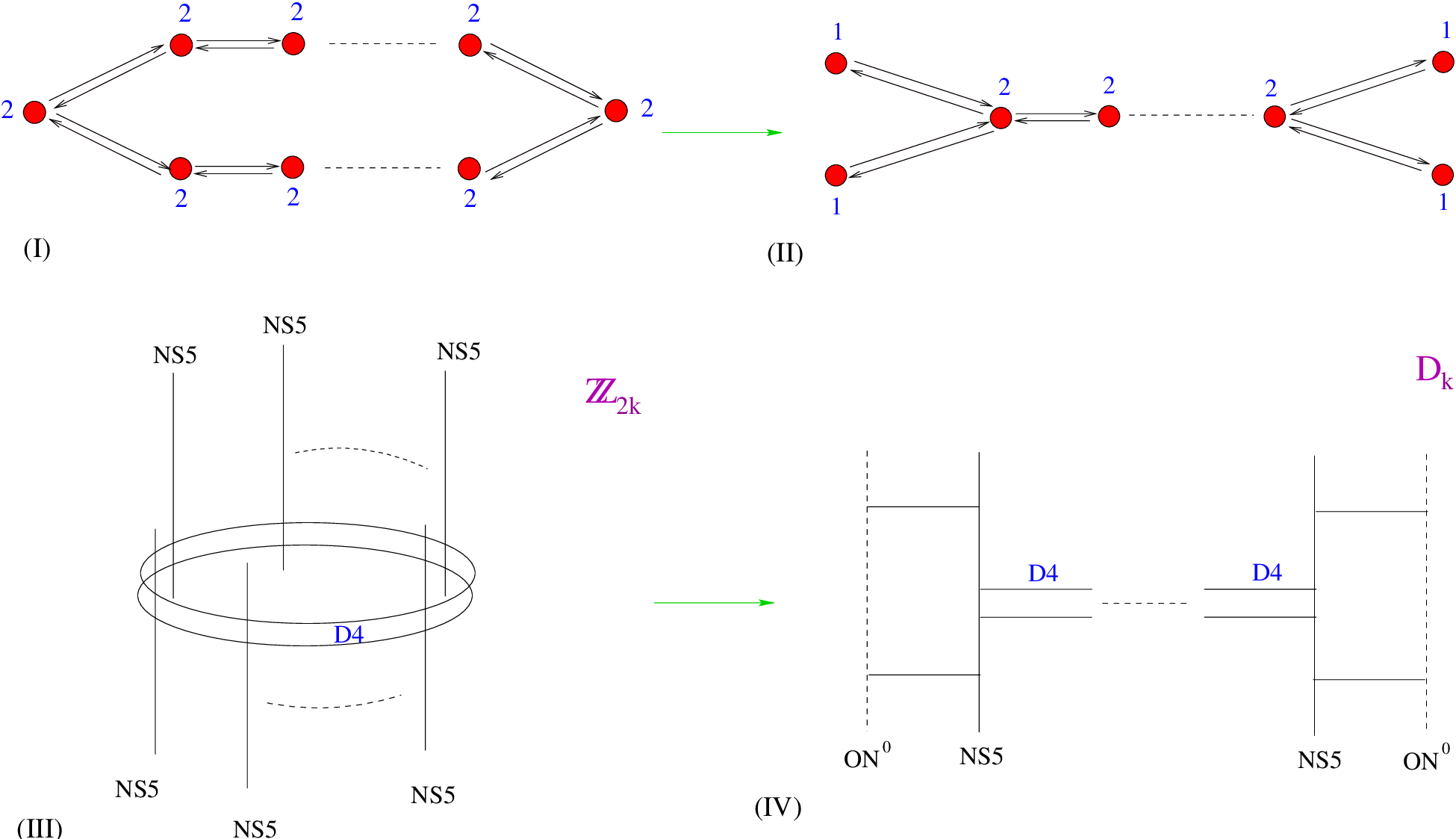,width=7in}
{
Obtaining the $D_k$ quiver (II) from the $\IZ_{2k}$ quiver (I) by the
stepwise projection algorithm. The brane setups are given respectively
in (IV) and (III).
\label{f:Dk}
}
\subsection{The $E_6$ Quiver from $D_2$}
We now move on to tackle the binary tetrahedral group $E_6$ (with the
relation that $E_6 / D_2 \cong \IZ_3$),
whose generators are
\[
\beta_4 = \left( \matrix{i & 0 \cr 0 & -i} \right), \quad 
\gamma=\left( \matrix{0 & i \cr i & 0} \right), \quad
\delta := {1 \over 2} \left( \matrix{ 1 - i & 1 - i  \cr -1 - i  & 
	1 + i  \cr  }\right).
\]
We observe therefore that it has yet one more generator $\delta$ than
$D_2$, hence we need to continue our stepwise projection from the
previous subsection, with the exception that we should begin with more
copies of $\IZ_4$. To see this let us first present
the irreducible matrix representations of the three generators of
$E_6$:
\[
\ba{c|c|c|c}
& \beta_4 & \gamma & \delta \\ \hline \hline
\Gamma^{(1)}_1 & 1 & 1 & 1 \\
\Gamma^{(1)}_2 & 1 & 1 & \omega_3 \\
\Gamma^{(1)}_3 & 1 & 1 & \omega_3^2 \\
\Gamma^{(2)}_4 & \beta_4 & \gamma & \delta \\
\Gamma^{(2)}_5 & \beta_4 & \gamma & \omega_3 \delta \\
\Gamma^{(2)}_6 & \beta_4 & \gamma & \omega_3^2 \delta \\
\Gamma^{(3)}_7 & \left( \matrix{ -1 & 0 & 0 \cr 0 & 1 & 0 \cr 0 & 0 &
		-1 \cr} \right) 
 		& \left(  \matrix{ 0 & 0 & -1 \cr 0 & -1 & 0 \cr -1 &
		0 & 0 \cr  } \right)
		& \left( \matrix{ -\frac{i}{2} & \frac{i}{\sqrt{2}} &  
     -\frac{i}{2} \cr -\frac{1}{\sqrt{2}} & 0 & \frac{1}
    {\sqrt{2}} \cr \frac{i}{2} & -\frac{i}{\sqrt{2}} & \frac{i}{2} \cr  } 
\right)
\ea
\]
The regular representation for these generators is therefore a matrix
of size $3 \cdot 1^2 + 3 \cdot 2^2 + 3^3 = 24$, in accordance with
\eref{reg}.

Our first step is as with the case of $D_2$, namely to change to a
convenient basis wherein $\beta_4$ becomes diagonal:
\beq
\label{betafinE6}
R^{reg}_{E_6}(\beta_4) = R^{reg}_{\IZ_4}(\beta_4) \otimes \I_6.
\eeq
The only difference between the above and \eref{betafinD2} is that we
have the tensor product with $\I_6$ instead of $\I_2$,
therefore at this stage we have a $\IZ_4$
quiver with the nodes labeled 6 as opposed to 2 as in Part (I) of
\fref{f:D2}. In other words we have 6 times the usual number of
D-brane probes.

Under the basis of \eref{betafinE6},
\beq
\label{gammafinE6}
R^{reg}_{E_6}(\gamma) = \left(
\matrix{\Sigma_3 & 0 & 0 & 0 \cr 0 & 0 & 0 & i\I_6 \cr
0 & 0 & \Sigma_3 & 0 \cr 0 & i\I_6 & 0 & 0 \cr}
\right)
\quad
{\rm where}
\quad
\Sigma_3 := \sigma_3 \otimes \I_3 = 
{\scriptsize
\left(\matrix{
1 & 0 & 0 & 0 & 0 & 0 \cr 0 & 1 & 0 & 0 & 0 & 0 \cr 0 & 0 & 1 & 0 & 0 & 
0 \cr 0 & 0 & 0 & -1 & 0 & 0 \cr 0 & 0 & 0 & 0 & 
-1 & 0 \cr 0 & 0 & 0 & 0 & 0 & -1 \cr}
\right).
}
\eeq
Subsequent projection gives a $D_2$ quiver as in part (II) of
\fref{f:D2}, but with the nodes labeled as $3,3,6,3,3$, three times
the usual. Note incidentally that \eref{betafinE6} and
\eref{gammafinE6} can be re-written in terms of regular
representations of $D_2$ directly: $R^{reg}_{E_6}(\beta_4) =
R^{reg}_{D_2}(\beta_4) \otimes \I_3$ and $R^{reg}_{E_6}(\gamma) =
R^{reg}_{D_2}(\gamma) \otimes \I_3$. To this fact we shall later turn.

To arrive at $E_6$, we proceed with one more projection,
by the last generator
$\delta$, the regular representation of which, observing the table
above, has the form (in the basis of \eref{betafinE6})
\beq
\label{deltafinE6}
R^{reg}_{E_6}(\delta) = \left(
\matrix{
S_1 & 0 & S_2 & 0 \cr 0 & \omega_8^{-1}P & 0 & \omega_8^{-1}P \cr
S_3 & 0 & S_4 & 0 \cr 0 & -\omega^8 P & 0 & \omega_8 P \cr}
\right)
\eeq
where
\[
S_1 := \left(\matrix{ 1 & 0 \cr 0 & 0
	\cr } \right) \otimes R^{reg}_{\IZ_3}(\beta_3), \quad
S_2 := \left(\matrix{ 0 & 0 \cr 1 & 0 \cr} \right) \otimes
	\left( \matrix{ 0 & 0 & 1 \cr 0 & 1 & 0 \cr 1 & 0 & 0 \cr  }
	\right),
\]
\[
S_3 := -i \left(\matrix{ 0 & 0 \cr 0 & 1 \cr} \right) \otimes
	\left( \matrix{ 0 & 0 & 1 \cr 0 & 1 & 0 \cr 1 & 0 & 0 \cr}
	\right),
\quad
S_4 := i \left(\matrix{ 0 & 1 \cr 0 & 0 \cr} \right) \otimes \I_3
\]
and
\[
P := R^{reg}_{\IZ_3}(\beta_3) \otimes \frac{1}{\sqrt{2}} \I_2;
\quad
\mbox{recalling that}
\quad
R^{reg}_{\IZ_3}(\beta_3) := \left(
\matrix{ 1 & 0 & 0 \cr 0 & \omega_3 & 0 \cr 0 & 0 & \omega_3^2 \cr  } 
\right).
\]
The inverse of \eref{deltafinE6} is readily determined to be
\[
R^{reg}_{E_6}(\delta)^{-1} = \left(
\matrix{
\tilde{S_1} & 0 & -S_3 & 0 \cr 0 & \frac12\omega_8 P^{-1} & 0 & 
	-\frac12\omega_8^{-1}P^{-1} \cr
S_2^T & 0 & -S_4^T & 0 \cr 0 & \frac12\omega_8 P^{-1} & 0 & 
	 \frac12\omega_8^{-1}P^{-1}\cr}
\right),
\qquad
\tilde{S_1} := \left(\matrix{ 1 & 0 \cr 0 & 0
	\cr } \right) \otimes R^{reg}_{\IZ_3}(\beta_3)^{-1}.
\]
Thus equipped, we must use \eref{proj} with \eref{deltafinE6} on the
matrix forms obtained in \eref{D2step2} (other fields can of course be
checked to have the same projection), with of course each number
therein now being $3 \times 3$ matrices.
The final matrix for $A^\mu$ is as in \eref{D2step2}, but with
\[
a_{11} = \left(
\matrix{a_{11(1)} & 0 & 0 \cr 0 & a_{11(2)} & 0 \cr 0 & 0 & a_{11(3)}}
\right)_{3 \times 3};
\quad
c_{11} = c_{22} = a_{22};
\quad
b = \left(
\matrix{b_{11} & 0 & 0 \cr 0 & b_{22} & 0 \cr 0 & 0 & b_{33} \cr }
\right)_{6 \times 6}
\]
where $a_{22}$, $c_{ii}$ are $3 \times 3$ while $b_{ii}$ are $2 \times
2$ blocks.
We observe therefore, that there are 7 distinct gauge group factors of
interest, namely $a_{11(1)}, a_{11(2)}, a_{11(3)}, a_{22}, b_{11},
b_{22}$ and $b_{33}$, with Dynkin labels $1,1,1,3,2,2,2$ respectively.
What we have
now is the $E_6$ quiver and the bifundamentals split and join
accordingly; the reader is referred to Part (I) of \fref{f:E6}.

\subsection{The $E_6$ Quiver from $\IZ_6$}
Let us make use of an interesting fact, that actually
$E_6 = \langle \beta_4, \gamma, \delta \rangle = \langle
\beta_4, \delta \rangle = \langle \gamma, \delta \rangle$. Therefore,
alternative to the previous subsection wherein we exploited the
sequence $\IZ_4 = \gen{\beta_4} \stackrel{+ \gamma}{\longrightarrow} D_2
\stackrel{+ \delta}{\longrightarrow} E_6$, we could equivalently apply
our stepwise projection on $\IZ_6 = \gen{\delta}
\stackrel{+ \beta_4}{\longrightarrow} E_6$. 

Let us first project with $\delta$, an element of order 6 and the
regular representation of which, after appropriate rotation is
\beq
\label{deltaE6}
R^{reg}_{E_6}(\delta) = R^{reg}_{\IZ_6}(\delta) \otimes \I_4.
\eeq
Therefore at this stage we have a $\IZ_6$ quiver with labels of
six 4's due to the $\I_4$; this is drawn in Part (II) of
\fref{f:E6}. The gauge group we shall denote as
$A^\mu := {\rm Diag}(a,b,c,d,e,f)_{24 \times 24}$, 
with $a, b, \cdots,f$ being $4 \times 4$ blocks.
\EPSFIGURE[ht]{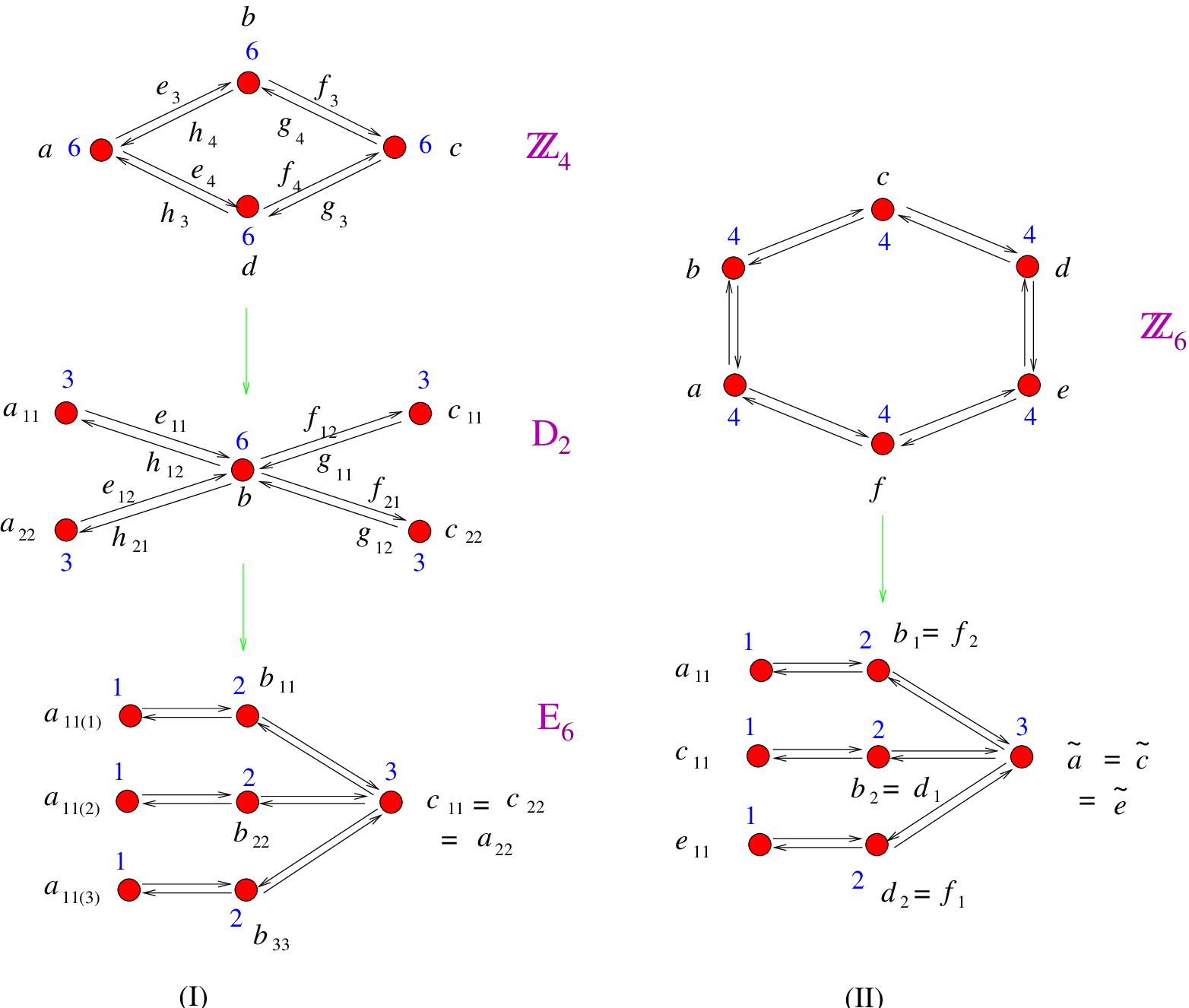,width=6.5in}
{
Obtaining the quiver diagram for the binary tetrahedral group $E_6$.
We compare the two alternative stepwise projections: (I) $\IZ_4 =
\gen{\beta_4} \rightarrow D_2 = \gen{\beta_4,\gamma} \rightarrow
E_6 = \gen{\beta_4,\gamma,\delta}$ and (II) $\IZ_6 = \gen{\delta}
\rightarrow E_6 = \gen{\delta,\beta_4}$.
\label{f:E6}
}

Next we perform projection by $R^{reg}_{E_6}(\beta_4)$
in the rotated basis, splitting and joining the gauge groups (nodes)
as follows
\[
A^\mu = 
{\scriptsize
\left(
\ba{cccccc}
\left(\matrix{a_{11} & 0 \cr 0 & \tilde{a}}\right) & 0 & 0 & 0 & 0 & 0 \\ 
0 & \left(\matrix{b_1 & 0 \cr 0 & b_2}\right) & 0 & 0 & 0 & 0 \\
0 & 0 & \left(\matrix{c_{11} & 0 \cr 0 & \tilde{c}}\right) & 0 & 0 & 0 \\ 
0 & 0 & 0 & \left(\matrix{d_1 & 0 \cr 0 & d_2} \right) & 0 & 0 \\
0 & 0 & 0 & 0 & \left(\matrix{e_{11} & 0 \cr 0 & \tilde{e}}\right) & 0 \\
0 & 0 & 0 & 0 & 0 & \left(\matrix{f_1 & 0 \cr 0 & f_2} \right)
\ea
\right)};
~~s.~t.~~
\ba{l}
\tilde{a} = \tilde{c} = \tilde{e},\\
b_2 = d_1,\\
d_2 = f_1,\\
f_2 = b_1,\\
\ea
\]
which upon substitution of the relations, gives us 7 independent factors:
$a_{11}, c_{11}$ and $e_{11}$ are numbers, giving 1
as Dynkin labels in the quiver; $b_1, b_2$ and
$d_2$ are $2 \times 2$ blocks, giving the 2 labels; while
$\tilde{a}$ is $3 \times 3$, giving the 3. We refer the reader to Part
(II) of \fref{f:E6} for the diagrammatical representation.
\section{Comments and Discussions}
Our procedure outlined above is originally inspired by a series of
papers \cite{Muto,Muto2,Muto3}, where the quivers for the $\Delta$
series of $\Gamma \subset SU(3)$ were observed to be obtainable from
the $\IZ_n \times \IZ_n$ series after an appropriate identification.
In particular, it was noted that 

$\Delta(3n^2) = \gen{\left\{ \IZ_n \times
\IZ_n := {\scriptsize \left( \matrix{\omega_n^i & 0 & 0 \cr 0 &
\omega_n^j & 0 \cr 0 
& 0 & \omega_n^{-i-j}}\right)_{i,j=0,\cdots,n-1} }\right\}, 
{\scriptsize \left(\matrix{ 0
& 0 & 1 \cr 1 & 0 & 0 \cr 
0 & 1 & 0 \cr}\right), \left(\matrix{ 0 & 1 & 0 \cr 0 & 0 & 1 \cr 1 &
0 & 0 \cr} \right)}}$ and subsequently the quiver for $\Delta(3n^2)$ is that of
$\IZ_n \times \IZ_n$ modded out by a certain $\IZ_3$ quotient. Similarly,
the quiver for
\[
\Delta(6n^2) = \langle \IZ_n \times \IZ_n, {\scriptsize
\left(\matrix{ 0 & 0 & 1 \cr 1 & 0 & 0 \cr 
0 & 1 & 0 \cr}\right), \left(\matrix{ 0 & 1 & 0 \cr 0 & 0 & 1 \cr 1 &
0 & 0 \cr} \right), \left(\matrix{ -1 & 0 & 0 \cr 0 & 0 & -1 \cr 0 &
-1 & 0 \cr  } \right), \left(  \matrix{ 0 & -1 & 0 \cr
-1 & 0 & 0 \cr 
0 & 0 & -1 \cr  }\right), \left( \matrix{ 0 & 0 & -1 \cr 0 & -1 & 0
\cr -1 & 0 & 0 \cr  }\right) } \rangle
\]
is that
of $\IZ_n \times \IZ_n$ modded out by a certain $S_3$ quotient.
In \cite{Muto3}, it was further commented that the $\Sigma$ series
could be likewise treated.

The motivation for those studies was to realise a brane-setup for the
non-Abelian $SU(3)$ orbifolds as geometrical quotients of the
well-known Abelian case of $\IZ_m \times \IZ_n$, viz., the Brane Box
Models. The key idea was to recognise that the irreducible
representations of these groups could be labelled by a double index
$(l_1,l_2) \in \IZ_n \times \IZ_n$ up to identifications.

Our purpose here is to establish an algorithmic treatment along
similar lines, which would be generalisable to arbitrary finite
groups. Indeed, since any finite group $\Gamma$ is finitely generated,
starting from the cyclic subgroup (with one single generator), our stepwise
projection would give the quiver for $\Gamma$ as appropriate splitting
and joining of nodes, i.e., as a certain geometrical action, of the
$\IZ_n$ quiver.
\subsection{A Mathematical Viewpoint}
To see why our stepwise projection works on a more axiomatic level, we
need to turn to a brief review of the Theory of Induced Representations.

It was a fundamental observation of Fr{\o}benius that
the representations of a group could be constructed from an arbitrary
subgroup. The aforementioned chain of groups, where we tried to relate
the regular representations, is precisely in this vein. Though we
shall largely follow the nomenclature of \cite{Leder}, we shall now
briefly review this theory in the spirit of the above discussions.

Let $\Gamma_1 = \gen{x_1,...,x_n}$ and $\Gamma_2 =
\gen{x_1,...,x_{n+1}}$. We see thus that $\Gamma_1 \subset
\Gamma_2$. Now let $R_{\Gamma_1}(x)$ be a representation (not
necessarily irreducible) of the element $x \in \Gamma_1$. Extending it to
$\Gamma_2$ gives
\[
R_{\Gamma_2}(y) = \left\{
\ba{ll}
R_{\Gamma_1}(x) & {\rm if} \quad y = x \in \Gamma_1 \\
0 & {\rm if} \quad y \not\in \Gamma_1
\ea
\right.
\]
It follows then that if we decompose $\Gamma_2$ as (right) cosets of
$\Gamma_1$,
\[
\Gamma_2 = \Gamma_1 t_1 \cup \Gamma_1 t_2 \cup \cdots \cup \Gamma_1 t_m
\]
we have an {\bf Induced Representation} of $\Gamma_2$ as
\beq
\label{induced}
R_{\Gamma_2}(y) = R_{\Gamma_1}(t_i y t_j^{-1}) =
\left(\ba{cccc}
R_{\Gamma_1}(t_1 y t_1^{-1}) & R_{\Gamma_1}(t_1 y t_2^{-1}) & \cdots &
	R_{\Gamma_1}(t_1 y t_m^{-1}) \\ 
R_{\Gamma_1}(t_2 y t_1^{-1}) & R_{\Gamma_1}(t_2 y t_2^{-1}) & \cdots &
	R_{\Gamma_1}(t_2 y t_m^{-1}) \\ 
\vdots & \vdots & & \vdots \\
R_{\Gamma_1}(t_m y t_1^{-1}) & R_{\Gamma_1}(t_m y t_2^{-1}) & \cdots &
	R_{\Gamma_1}(t_m y t_m^{-1}) 
\ea\right).
\eeq
A beautiful property of \eref{induced} is that it has only one member
of each row or column non-zero and whereby it is essentially a
generalised permutation (see e.g., 3.1 of \cite{Leder}) matrix acting
on the $\Gamma_1$-stable submodules of the $\Gamma_2$-module.

Now, for the case at hand the coset decomposition is simple due to the
addition of a single new generator: the (right) transversals
$t_1, \cdots, t_m$ are simply powers of the extra generator $x_{n+1}$
and $m$ is simply the index of $\Gamma_1 \subset \Gamma_2$, namely
$|\Gamma_2|/|\Gamma_1|$, i.e.,
\beq
\label{ti}
t_i = x_{n+1}^{i-1} \qquad i = 1,2,\cdots,m; \quad 
	m=\frac{|\Gamma_2|}{|\Gamma_1|}.
\eeq

Now let us define an important concept for an element $x \in \Gamma_2$
\begin{definition}
We call a representation $R_{\Gamma_2}(x)$ {\bf factorisable} if it
can be written, up to possible change of bases, as a tensor product
$R_{\Gamma_2}(x) = R_{\Gamma_1}(x) \otimes \I_k$ for some integer $k$.
\end{definition}
Factorisability of the element, in the physical
sense, corresponds to the ability to initialise our stepwise
projection algorithm, by which we mean that the orbifold projection by
this element is performed on $k$ copies as in the usual sense, i.e.,
a stack of $k$ copies of the quiver. Subsequently we could continue
with the stepwise algorithm to demonstrate how the nodes of these
copies merge or split. In the corresponding D-brane picture this simply
means that we should consider $k$ copies of each image
D-brane probe in the covering space.

The natural question to ask is of course why our examples in the
previous section permitted factorisable generators so as to in turn
permit the performance of the stepwise projection.
The following claim shall be of great assurance to us:
\begin{proposition}
Let $H$ be a subgroup of $G$, then the representation $R_G(x)$ for an
element $x \in H \subset G$ induced
from $R_H(x)$ according to \eref{induced} is factorisable and $k$ is
equal to $|G|/|H|$, the index of $H$ in $G$.
\end{proposition}
Proof: 
Take $R_H(x \in H)$, and tensor it with $\I_{k=|G|/|H|}$; this remains
of course a representation for $x\in H$. It then remains to find the
representations of $x \not\in H$, which we supplement by the
permutation actions of these elements on the $H$-cosets. At the end of
the day we arrive at a representation $R'_G(x)$ of dimension $k$, such
that it is factorisable for $x\in H$
and a general permutation for $x \not\in H$. However by the uniqueness
theorem of induced representations (q.v. e.g. \cite{Serre} Thm 11)
such a linear representation $R'_G(x)$ must in fact be
isomorphic to $R_G(x)$. Thus by
explicit construction we have shown that $R_G(x\in H) = R_H(x) \otimes
\I_k$. $\stackrel{~}{\tiny \sq}$

We can be more specific and apply Proposition 4.1 to our case of the
two groups the second of which is generated by the first with one
additional generator. Using the elegant property that the induction
of a regular representation remains regular (q.v. e.g., 3.3 of
\cite{Serre}), we have:
\begin{corollary}
Let $\Gamma_1$ and $\Gamma_2$ be as defined above, then
\[
R^{reg}_{\Gamma_2}(x_i) = R^{reg}_{\Gamma_1}(x_i) \otimes
\I_{|\Gamma_2|/|\Gamma_1|} \qquad \mbox{for common generators} \qquad
i = 1,2,\ldots,n.
\]
\end{corollary}
In particular, since any $G = \gen{x_1,\ldots,x_n}$ contains a cyclic
subgroup generated by, say $x_1$ of order $m$, i.e., $\IZ_m =
\gen{x_1}$, we conclude that
\begin{corollary}
$R^{reg}_G(x_1) = R^{reg}_{\IZ_m}(x_1) \otimes \I_{|G|/m}$, and hence
the quiver for $G$ can always be obtained by starting with the $\IZ_m$
quiver using the stepwise projection.
\end{corollary}

Let us revisit the examples in the previous section equipped with the
above knowledge.
For the case of $\Gamma_1 = \IZ_4 = \gen{\beta_4}$ and $\Gamma_2
= D_2$ with the extra generator $\gamma$, \eref{ti} becomes 
$t_1 = \I$ and $t_2 = \gamma$ as the index of $\IZ_4$ in $D_2$ is
$\frac{|D_2|=8}{|\IZ_4|=4}=2$.
The induced representation of $\beta_4$ according to \eref{induced}
reads
\[
R_{D_2}(\beta_4) =
\left(
\matrix{R^{reg}_{\IZ_4}(\I\beta_4\I^{-1}) &
		R^{reg}_{\IZ_4}(\I\beta_4\gamma^{-1}) \cr 
	R^{reg}_{\IZ_4}(\gamma\beta_4\I^{-1}) &
		R^{reg}_{\IZ_4}(\gamma\beta_4\gamma^{-1} )}
\right)
=
\left(
\matrix{R^{reg}_{\IZ_4}(\beta_4) & 0 \cr
	0 & R^{reg}_{\IZ_4}(\beta_4^{-1})}
\right)
\]
using the fact that $\gamma \beta_k \gamma^{-1} = \beta_k^{-1}$ in
$D_k$ for the last entry.  Recalling that $R^{reg}_{\IZ_4}(\beta_4)
= {\scriptsize \left( \matrix{ 1 & 0 & 0 & 0 \cr 0 & i & 0 & 0 \cr
0 & 0 & i^2 & 0 \cr 0 & 0 & 0 & i^3 \cr  }\right)}$, this is
subsequently equal to  
$R^{reg}_{\IZ_4} \otimes \I_2$ after appropriate permutation of basis.
Thus Corollary 4.1 manifests her validity as we see that the $R_{D_2}$
obtained by Fr{\o}benius induction of $R^{reg}_{\IZ_4}$ is indeed
regular and moreover factorisable, as \eref{betafinD2} dictates.

Similarly with the case of $\IZ_6 \rightarrow E_6$, we see that
Corollary 4.1 demands that for the common generator $\delta$,
$R^{reg}_{E_6}(\delta)$ should be factorisable, as is indeed
indicated by \eref{deltaE6}. So too is it with $\IZ_4 \rightarrow E_6$,
where $R^{reg}_{E_6}(\beta_4)$ should factorise, precisely as shown by
\eref{betafinE6}.

The above have actually been special cases of Corollary 4.2, where we
started with a cyclic subgroup; in fact we have also presented an
example demonstrating the general truism of Proposition 4.1. In the
case of $D_2 \rightarrow E_6$, we mentioned earlier that
$R^{reg}_{E_6}(\beta_4) =
R^{reg}_{D_2}(\beta_4) \otimes \I_3$ and $R^{reg}_{E_6}(\gamma) =
R^{reg}_{D_2}(\gamma) \otimes \I_3$ for the common generators as was
seen from \eref{betafinE6} and \eref{gammafinE6}; this is exactly as
expected by the Proposition.
\subsection{A Physical Viewpoint: Brane Setups?}
Now mathematically it is clear what is happening to the quiver as we
apply stepwise projection. However this is only half of the story; as
we mentioned in the introduction, we expect T-duality to take D-branes
at generic orbifold singularities to brane setups.
It is a well-known fact that the brane setups for the $A$ and $D$-type
orbifolds $\C^2/\IZ_n$ and $\C^2/D_n$ have been realised (see
\cite{Han-Zaf,Han-Ura} and \cite{Kapustin} respectively). It has been the main
intent of a collective 
of works (e.g \cite{Bo-Han,Muto2,Muto3}) to establish such setups for
the generic singularity. 

In particular, the problem of finding a consistent brane-setup for the
remaining case of the exceptional groups
$E_{6,7,8}$ of the $ADE$ orbifold singularities of $\C^2$ (and indeed
analogues thereof for $SU(3)$ and $SU(4)$ subgroups) so far has been
proven to be stubbornly intractable. An original motivation for the
present work is to attempt to formulate an algorithmic outlook wherein
such a problem, with the insight of the algebraic structure of an
appropriate chain of certain relevant groups, may be addressed
systematically.
\subsubsection{The $\IZ_2$ Action on the Brane Setup}
Let us attempt to recast our discussion in Subsection 3.1 into a
physical language. First we try to interpret the action by
$R^{reg}_{D_k}(\gamma)$ in \eref{gammafin} on the $\IZ_{2k}$ quiver as
a string-theoretic action on brane setups to get the corresponding
brane setup of $D_k$ from that of $\IZ_{2k}$.

Now the brane configuration for the $\IZ_{2k}$ orbifold is the
well-known {\em elliptic model} consisting of $2k$ NS5-branes arranged
in a circle with D4-branes stretched in between as shown in Part (III)
of \fref{f:D2}. After stepwise projection by $\gamma$, the quiver in
Part (I) becomes that in Part(II) (see \fref{f:Dk} also). There is an
obvious $\IZ_2$ quotienting involved, where the nodes $i$ and $2k-i$
for $i=1,2,...,k-1$ are identified while each of the nodes $0$ and $k$ 
splits into two parts. Of course, this symmetry is not immediately
apparent from the properties of $\gamma$, which is a group element of
order 4. This phenomenon is true in general: {\sf the order of the
generator used in the stepwise projection does not necessarily
determine what symmetry the parent quiver undergoes to arrive at the
resulting quiver; instead we must observe {\it a posteriori} the shapes of the
respective quivers.}

Let us digress a moment to formulate the above results in the language
used in \cite{Muto,Muto2}. Recalling from the brief comments in the
beginning of Section 4, we adopt their idea of labelling 
the irreducible representations of $\Delta$ by $\IZ_n \times \IZ_n$ up
to appropriate identifications, which in our terminology is simply the
by-now familiar stepwise projection of the parent $\IZ_n \times \IZ_n$ quiver.
As a comparison, we apply this idea to the case of $\IZ_{2k}
\rightarrow D_k$. Therefore we need to label the irreps of $D_k$ or
appropriate tensor sums thereof, in terms of certain (reducible)
2-dimensional representations of $\IZ_{2k}$.
Motivated by the factorization property \eref{betafinD2}, we chose
these representations to be
\beq
\label{z2kindex}
R_{\IZ_{2k}(2)}^l := R_{\IZ_{2k}(1)}^{l,irrep} 
\oplus R_{\IZ_{2k}(1)}^{l,irrep}
\eeq
where $l \in \IZ_{2k}$, and amounts to precisely a $\IZ_{2k}$-valued index
on the representations of $D_k$ 
(since $\IZ_{2k}$ is Abelian), which with foresight, we shall later use
on $D_k$.
We observe that such a labelling scheme has a symmetry
\[
R_{\IZ_{2k}(2)}^l \cong R_{\IZ_{2k}(2)}^{-l},
\]
which is obviously a $\IZ_2$ action. Note that $l=0$ and $l=k$ are
fixed points of this $\IZ_2$.

We can now associate the 2-dimensional irreps of $D_k$ with the non-trivial
equivalence classes of the $\IZ_{2k}$ representations \eref{z2kindex},
i.e., for $l=1,2,\ldots,k-1$ we have 
\beq
R_{\IZ_{2k}(2)}^l \cong R_{\IZ_{2k}(2)}^{-l} 
\rightarrow R_{D_{k}(2)}^{l,irrep}. 
\label{map1}
\eeq
These identifications correspond
to the merging nodes in the associated quiver diagram. 
As for the fixed points, we need to map
\beq
\ba{l}
R_{\IZ_{2k}(2)}^0 \rightarrow R_{D_{k}(1)}^{1,irrep} 
\oplus R_{D_{k}(1)}^{2,irrep} \\
R_{\IZ_{2k}(2)}^k \rightarrow R_{D_{k}(1)}^{3,irrep} 
\oplus R_{D_{k}(1)}^{4,irrep}.
\ea
\label{map2}
\eeq 
These fixed points are associated precisely with the nodes that split.

This construction shows clearly how, in the labelling scheme of
\cite{Muto,Muto2}, our stepwise algorithm derives
the $D_k$ quiver as a $\IZ_2$ projection of the $\IZ_{2k}$ quiver. 
The consistency
of this description is verified by substituting the 
representations $R_{\IZ_{2k}(2)}^l$ in the $\IZ_{2k}$ quiver 
relations ${\cal R} \otimes R_{\IZ_{2k}(2)}^l = \bigoplus\limits_{\bar l}
a_{l {\bar l}}^{\IZ_{2k} ({\cal R})}  R_{\IZ_{2k}(2)}^{\bar l}$ using
\eref{map1} and \eref{map2}, which results 
exactly in the $D_k$ quiver relations. 
We can of course apply the stepwise projection for the case of 
$\IZ_n \times \IZ_n \rightarrow \Delta$, and would arrive at the results
in \cite{Muto,Muto2}.

In the brane setup picture, the identification of the
nodes $i$ and $2k-i$ for $i=1,2,...,k-1$ corresponds to the
identification of these intervals of NS5-branes as well as the
D4-branes in between in the $X^{6789}$
directions (with direction-6 compact). Thus the $\IZ_2$ action on the
$\IZ_{2k}$ quiver should include a space-time action which identifies
$X^{6789}=-X^{6789}$. Similarly, the splitting of gauge fields in
intervals $0$ and $k$ hints the existence of a $\IZ_2$ action on the
string world-sheet. Thus the overall $\IZ_2$ action should include two
parts: a space-time symmetry which identifies and a world-sheet
symmetry which splits respective gauge groups.

What then is this action physically? What object in string theory
performs the tasks in the above paragraph? Fortunately, the
space-time parity and string world-sheet $(-1)^{F_L}$ actions
\cite{Sen,Kapustin} are precisely the aforementioned symmetries. In
other words, the {\em ON-plane} is that which we seek.
This is of great assurance to us, because the brane setup for $D_k$
theories, as given in \cite{Kapustin}, is indeed a configuration which
uses the ON-plane to project out or identify fields in a manner
consistent with our discussions.
\subsubsection{The General Action on the Brane Setup?}
It seems therefore, that we could now be boosted with much confidence:
since we have proven in the previous subsection that our stepwise
projection algorithm is a constructive method of arriving at {\em any}
orbifold quiver by appropriate quotient of the $\IZ_n$ quiver, could we
not simply find the appropriate object in string theory which would
perform such a quotient, much in the spirit of the orientifold
prescribing $\IZ_2$ in the above example, on the well-known $\IZ_n$
brane setup, in order to solve our problem?

Such a confidence, as is with most in life, is overly optimistic. Let
us pause a moment to consider the $E_6$ example. The action by
$\delta$ in the case of $D_2 \rightarrow E_6$ in \S 3.2 and
that of $\beta_4$ in the case of $\IZ_6 \rightarrow E_6$ in \S 3.3
can be visualised in Parts (I) 
and (II) of \fref{f:E6} to be an $\IZ_3$ action on the respective
parent quivers. In particular, the identifications $c_{11} \sim c_{22}
\sim a_{22}$ and $\tilde{a} \sim \tilde{c} \sim \tilde{e}; b_1 \sim
f_2, b_2 \sim d_1, d_2 \sim f_1$ respectively for Parts (I) and (II) 
are suggestive of a $\IZ_3$ action on $X^{6789}$. The tripartite
splittings for $b, a_{11}$ and $a,b,d$ respectively also hint at a
$\IZ_3$ action on the string world-sheet.

Again let us phrase the above results in the scheme of
\cite{Muto,Muto2}, and manifestly show how
the $E_6$ quiver results from a $\IZ_3$ projection of the $D_2$ quiver.
We define the following representations of $D_2$: 
$R_{D_{2}(6)}^0 = R_{D_{2}(2)}^{irrep} \oplus  R_{D_{2}(2)}^{irrep}
\oplus  R_{D_{2}(2)}^{irrep}$ and $R_{D_{2}(3)}^l = 
R_{D_{2}(1)}^{l,irrep}  \oplus R_{D_{2}(1)}^{l,irrep}  \oplus
R_{D_{2}(1)}^{l,irrep}$ where $l \in \IZ_4$ labels the four 1-dimensional
irreducible representations of $D_2$. There is an identification
\[
R_{D_{2}}^l \cong  R_{D_{2}}^{f(l)}
\]
where
\[
f(l) = \left\{\begin{array}{c}
0, \;\; l=0 \\
2, \;\; l=1 \\
3, \;\; l=2 \\
1, \;\; l=3 \\
\end{array}\right.
\]
Clearly this is a $\IZ_3$ action on the index $l$.
Note that we have two representations labelled with $l=0$ 
which are fixed points of this action. 
In the quiver diagram of $D_2$ these correspond to the middle node
and another one arbitrarily selected from the
remaining four, both of which split into three. 
The remaining three nodes are consequently merged into a single one 
(see \fref{f:E6}). 
To derive the $E_6$ quiver we need to map the nodes of the parent $D_2$
quiver as
\[
\ba{c}
R_{D_{2}(6)}^0 \rightarrow 
R_{E_{6}(2)}^{1,irrep} \oplus R_{E_{6}(2)}^{2,irrep} 
\oplus R_{E_{6}(2)}^{3,irrep}\\
R_{D_{2}(3)}^0 \rightarrow 
R_{E_{6}(1)}^{1,irrep} \oplus R_{E_{6}(1)}^{2,irrep} 
\oplus R_{E_{6}(1)}^{3,irrep}\\
R_{D_{2}(3)}^l \cong R_{D_{2}(3)}^{f(l)} \rightarrow  R_{E_{6}(3)}^{irrep},
\qquad l \in
\IZ_{4}-\{0\}. \\
\ea
\]
Consistency requires that if we replace $R_{D_{2}}$ in the $D_2$ quiver 
defining relations and then use the above mappings, we get the $E_6$ quiver
relations for $R_{E_{6}}^{irrep}$.

The origin of this $\IZ_3$ analogue of the orientifold
$\IZ_2$-projection is
thus far unknown to us. If an object with this property is to exist,
then the brane setup for the $E_6$ theory could be implemented; on the
other hand if it does not, then we would be suggested at why the
attempt for $E_6$ has been prohibitively difficult.

The $\IZ_3$ action has been noted to arise in \cite{Muto2} in the
context of quotienting the $\IZ_n \times \IZ_n$ quiver to arrive at the
quiver for the $\Delta$-series. Indeed from our comparative study in Section
4.2.1, we see that in general, labelling the irreps by a
multi-index is precisely our stepwise algorithm in disguise, as
applied to a product Abelian group: the $\IZ_n \times \cdots \times
\IZ_n$ orbifold. Therefore in a sense we have explained why the
labelling scheme of \cite{Muto,Muto2} should work.

And the same goes with $E_7$ and $E_8$: we could perform stepwise
projection thereupon and mathematically obtain their quivers as
appropriate quotients of the $\IZ_n$ quiver by the symmetry $S$ of the
identification and splitting of nodes. To find a physical brane setup,
we would then need to find an object in string theory which has an
$S$ action on space-time and the string world-sheet. Note that the above
are cases of the $\C^2$ orbifolds; for the $\C^k$-orbifold we should
initialise our algorithm with, and perform stepwise projection on the
quiver of $\IZ_n \times \cdots \times \IZ_n$ ($k-1$ times), i.e., the
brane box and cube ($k=2,3$).

Though mathematically we have found a systematic treatment of
constructing quivers under a new light, namely the ``stepwise
projection'' from the Abelian quiver, much work remains. In the field
of brane setups for singularities, our algorithm is intended to be a
small step for an old standing problem. We must now diligently seek
a generalisation of the orientifold plane with symmetry $S$ in
string theory, that could perform the physical task which our
mathematical methodology demands.
\section*{Acknowledgements}
{\it Ad Catharinae Sanctae Alexandriae et Ad Majorem Dei Gloriam...\\}
We would like to extend our gratitude to D. Berenstein for useful 
discussions, especially for his informing us of his related
works in the context of discrete torsion. Furthermore, we
are indebted to the Reed Fund, the CTP, and the LNS for
their gracious patronage.

\bibliographystyle{JHEP}

\end{document}